# ALFRED: A Methodology to Enable Component Fault Trees for Layered Architectures

Kai Höfig, Marc Zeller, Reiner Heilmann

**Abstract**—Identifying drawbacks or insufficiencies in terms of safety is important also in early development stages of safety critical systems. In industry, development artefacts such as components or units, are often reused from existing artefacts to save time and costs. When development artefacts are *reused*, their existing safety analysis models are an important input for an early safety assessment for the new system, since they already provide a valid model. Component fault trees support such reuse strategies by a compositional horizontal approach. But current development strategies do not only divide systems *horizontally*, e.g., by encapsulating different functionality into separate components and hierarchies of components, but also vertically, e.g. into software and hardware architecture layers. Current safety analysis methodologies, such as component fault trees, do not support such vertical layers.

Therefore, we present here a methodology that is able to divide safety analysis models into different layers of a systems architecture. We use so called *Architecture Layer Failure Dependencies* to enable component fault trees on different layers of an architecture. These dependencies are then used to generate safety evidence for the entire system and over all different architecture layers. A case study applies the approach to hardware and software layers.

✦

## 1 INTRODUCTION

The development of todays safety or mission critical systems underlies a series of legislative and normative regulations making safety and reliability the most important non-functional properties of embedded systems in many application domains of embedded systems, such as aerospace, railway, health care, automotive and industrial automation. Thus, along with the growing system complexity, also the need for safety assessment as well as its effort is increasing drastically in order to guarantee the high quality demands in these application domains. However, this trend is contrary to industry's aim to reduce development costs and time-to-market of new products. Particularly in the case of software intensive embedded systems, their complexity is rapidly increasing and extended analysis techniques are required that scale to the **increasing system complexity**.

• K. Höfig, M. Zeller, R. Heilmann are with Siemens AG, Corporate Technology, Munich, Germany.
 E-mail: {firstname}.{lastname}@siemens.com

The rapid development and analysis of safety analysis models is important also in early stages during the development of safety critical systems. Such models aim at identifying drawbacks or insufficiencies in terms of safety. The early identification of such drawbacks is crucial for a cost efficient development process. In industry, development artefacts such as components or units, are often reused from existing artefacts to save time and costs and increase product quality.

Changes are made to these artefacts to match the requirements for a new product. In software development such a reuse strategy is also known as *clone and own*. When development artefacts are *reused*, their existing safety analysis models are an important input for an early safety assessment for a new system or product, since they already provide a valid model in terms of the former system or product. Nevertheless, changes and adoptions during the new development invalidate former analyses and require an adoption to the changes.

Another area where the rapid and automated development of safety analysis models is es-sential are safety critical cyberphysical systems. Cyberphysical systems consist of more or less loosely coupled embedded systems. Their exact alignment is unclear at design time and the possible configurations at runtime are typically infinite. So, each embedded system, as a part of a cyberphysical system, is *reused* in many different configurations. For a safety critical function of such systems it might be necessary to be certified automatically *at runtime* to assure a safe operation.

For fault tree analysis, models exist that support such reuse strategies by a compositional strategy. Fault tree elements are related to their development artefacts and can be reused along with the reused development artefact. Modular or compositional safety analysis methodologies such as component fault trees [1] or HipHops [2] ease the adoption of changes for existing development artefacts by constraining the adoption activities for safety to the artefacts that require changes and provide benefits for an automated proof.

But current development strategies do not only divide systems *horizontally*, e.g., by encapsulating different functionality into separate components, but also vertically. An example for a vertical decomposition of a system is the distinction between software and hardware. Another example is the decomposition into a functional layer and a physical layer. Horizontal and vertical decompositions reduce complexity and increase reusability by applying the principles of *divide and conquer* and *separation of concerns*. As can be seen in section 2, current safety analysis methodologies allow a horizontal decomposition of safety analysis models, even into arbitrary levels of hierarchy, but lack at dividing models into different vertical layers and then generate safety evidence from it.

Therefore, we present in this paper a methodology that is able to divide safety analysis models into different vertical layers of a sys tems architecture. We use so called *Architecture Layer FailuRE Dependencies* (ALFRED) to enable component fault trees on different layers of an architecture. The architecture layer failure dependencies are then used to generate safety evidence for the entire system and over all different architectural layers maintaining important properties of a vertically layered architecture model, such as transparency and independency.

The rest of this document is structured as follows: first the related approaches are summarized in section 2. In section 3, the methodology of component fault trees is described. Section 4 presents the central methodology of this paper and describes the application using a small example. Section 5 applies the methodology to a compact system from the automotive domain. Section 6 summarizes this paper and provides a prospect for future work.

## 2 RELATED WORK

The research presented in this paper is related to the general research area on model-based safety evaluation of software system and reuse of safety artefact as a sub-area of model-based safety evaluation.

**Model-based Safety Evaluation.** The use of models in safety engineering processes has gained increasing attention in the last decade [3], [4], [5]. Specifically the idea is to support automatic generation of safety artifacts such as fault trees [6], [7], [8], [9], [10], [11], [12], [13], [14], [15], [16], [17], [18], [19], [20], [21] or FMEA tables [22], [23], [24], [25], [26], [27] from system models. To construct the safety artefact, the system models are often annotated with failure propagation models [28], [29], [11], [1], [30], [15]. These failure propagation models are commonly combinatorial in nature thus producing static fault trees. This is also driven by the industrial need to certify [31], [32], [33] their system with static fault trees. Only rarely more advanced safety evaluation models such as Dynamic Fault Trees (DFTs) [34], [35], [36], [37], Generalized Stochastic Petri Nets (GSPNs)[38], State-Event Fault Trees (SEFTs) [39], [30], [40] or Markov models exist [41], [42]. Beside annotating an architecture specification, there are also approaches to construct a safety artefact via model checking techniques [43], [44], [45], [46], [47].

The approach presented in this paper complements these techniques, especially those one that generate fault trees from component-based failure propagation models. For these model-based safety evaluation approaches, including vertically decomposed architectures could add significant productivity gains.

**Reusable Model-based Safety Analysis.** In the development of a safety-critical system similar components may result in similar safety evaluation models. However, significant reuse of safety evaluation models is still rare and should be handled with great care. In fault trees a common way of reusing is to copy an entire subtree. However, Kaiser et al. [39], [1], [30], [40] realized that fault trees should be reused as subgraphs, and created Component Fault Trees and State-Event Fault Trees (SEFTs). The line of research results the ability to reuse an entire subgraph, via typing them and so called in- and out-failure ports that serve as interfaces to the rest of the fault tree. In another line of research Wolforth et al. [48], [49], [50] created a language extension for HiP-HOPS [15] that allow for pattern-based specification of reusable failure propagation models. As a result, reuse of safety models between similar components and from common patterns is possible. From the reused failure propagation models via the HiP-HOPS [15] methodology complete fault trees can be automatically constructed.

Our approach also aims at reuse, but differs from the presented approaches because we are focusing on a vertically decomposed architecture.

In the next section, the concept of component fault trees is introduced and later extended in section 4 for a vertical decomposed architecture model.

## 3 COMPONENT FAULT TREES

A component fault tree is a Boolean model associated to system development elements such as components [1]. It has the same expressive power as classic fault trees that can for example be found in [51]. As classic fault trees, also component fault trees are used to model failure behavior of safety critical systems. This failure behavior is used to document that a system is safe and can also be used to identify drawbacks of the design of a system.

A separate *component fault tree element* is related to a component. Failures that are visible at the outport of a component are models using *Output Failure Modes* which are related to the specific outport. To model how specific failures propagate from an inport of a component to the outport, *Input Failure Modes* are used. The inner failure behavior that also influences the output failure modes is modeled using the gates *NOT*, *AND*, *OR*, and *Basic Event*.

Every component fault tree can be transformed to a classic fault tree by removing the input and output failure modes elements. Figure 1 shows on the left side a classic fault tree and on the right side a component fault tree. In both trees, the topevents or output events *TE1* and *TE2* are modeled. The component fault tree model allows, additionally to the Boolean formulae that are also modeled within the classic fault tree, to associate the specific topevents to the corresponding ports where these failures can appear. Topevent *TE1* for example appears at port *O1*. Using this methodology of components also within fault tree models, benefits during the development can be observed, for example an increased maintainability of the safety analysis model [52].

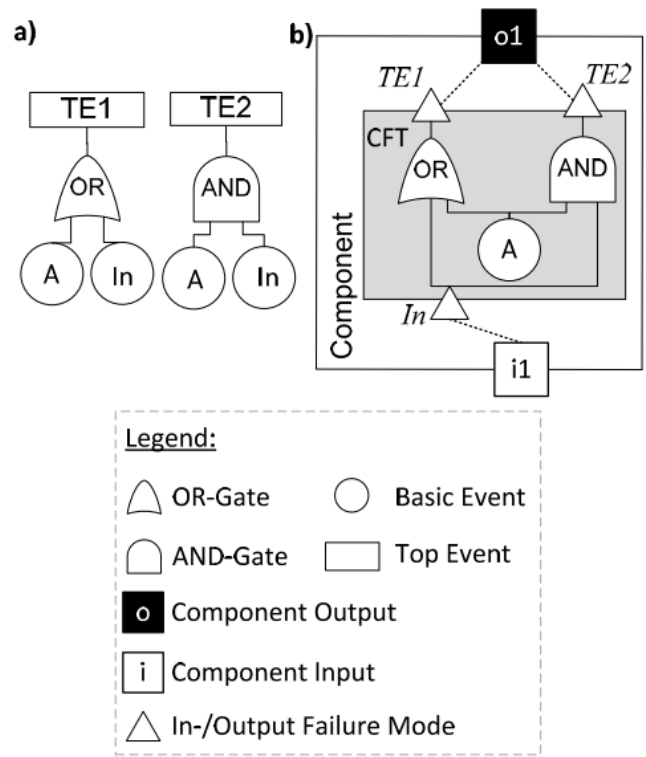

Fig. 1. Classic Fault Tree (a) and Component Fault Tree (b) [52].

Component fault trees and also comparable approaches as mentioned in section 2, require elements modeling information flow from one component to another. In figure 1, the ports *o1* and *i1* are examples of such elements. On the one hand, this is beneficial for a compositional approach since it makes the component and its related safety model exchangeable to a new context. On the other hand, such approaches require need to include *all* dependencies using ports or similar elements to allow an analysis on system level. Dependencies in the safety model to layers that are modeled independently and without such port elements are not intended. Adding such port elements to the architecture is not an acceptable solution, since it would invalidate the benefits of having vertical abstractions using independent layers in the architecture model.

In the next section, the approach of component fault trees is extended to overcome this drawback.

## 4 ARCHITECTURE LAYER FAILURE DEPENDENCIES

In this section, we describe how architecture layer failure dependencies can be used to divide a safety analysis model based on component fault trees into different layers and still maintain an analysis for an entire system without the need to model the information flow between layers using explicit port elements as described in section 3.

Figure 2 shows an example system with failure dependency relations between different layers of the system model. On the top, the component fault trees (CFT) of two functions $f_1$ and $f_2$ are depicted. Function $f_1$ receives two sensor values at its in ports. If both values are unavailable, the result of the function is unavailable (AND gate). If function $f_2$ receives no signal from function $f_1$, function $f_2$ is not available (loss off). These two functions build one layer of the architecture model, e.g. the software or functional layer.

The second layer consists of two hardware or physical components. Figure 2 shows the component fault trees for these components: $RAM$ and $CPU$. These components represent the memory (RAM) and computational resource (CPU) of the system. These components are in a failure mode (loss of) if either basic event $a$ occurs in the CPU the or if basic event $b$ occurs in the memory component.

Since the *functional* failure behavior is also dependent from failures that occur in the *hardware* layer, failure dependency relations are used here to model this dependency. It is assumed, that function $f_1$ can only be executed if both, memory and computational power is available. This is reflected by the dependency relations from the component fault tree for function $f_1$ to both component fault tree elements of the physical layer, $CPU$ and $RAM$. Function $f_2$ is only dependent of the correct function of the memory resource and so the failure dependency is only related to the $RAM$ CFT in the physical layer. In the following we formalize such systems.

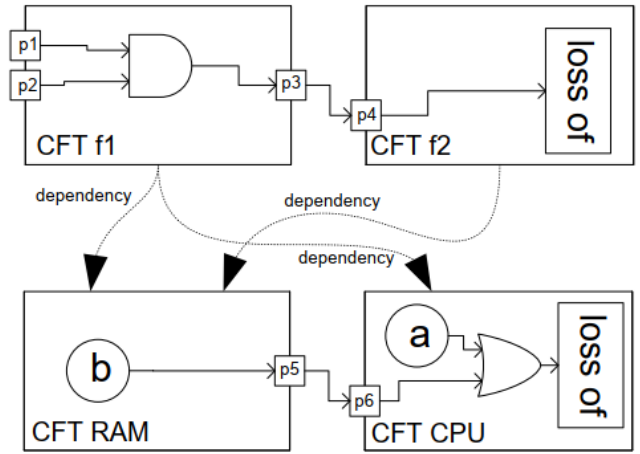

Fig. 2. Example system with component fault trees, two different layers (hardware and software) and failure dependency relations (dotted lines).

So, let $C = c_1, \ldots, c_n$ be a set of components of a system, $CFT = cft_1, \ldots, cft_m$ the set of component fault trees of the corresponding components with

$$\tilde{CFT}(c) = cft \text{ with } c \in C \text{ and } cft \in CFT.$$

Let

$$I(c) = in_1, \ldots, in_i, \text{ and}$$

$$O(c) = out_1, \ldots, out_j$$

be the in- and outports of component $c$. Let

$$\overline{PC} = \{(out, in) \mid out \in O(c_1) \cup \ldots \cup O(c_n),\\ in \in I(c_1) \cup \ldots \cup I(c_n)\}$$

be the set of all possible port connections and

$$PC \subseteq \overline{PC}$$

be the set of actual port connections modeling the data flow from the outport of a component to the inport of another component.

Let

$$A(c) = \{x | x = C\tilde{F}T(d), d \in CFT\}$$

be the set of all failure dependencies of component $c$ to other components.

For the example system, the previously defined sets are as follows:

$$\begin{aligned} C &= f_1, f_2, RAM, CPU \\ I(f_1) &= p_1, p_2 \\ I(f_2) &= p_4 \\ I(RAM) &= \{\} \\ I(CPU) &= p_6 \\ O(f_1) &= p_3 \\ O(f_2) &= \{\} \\ O(RAM) &= p_5 \\ O(CPU) &= \{\} \\ PC &= (p_3, p_4), (p_5, p_6) \\ A(f_1) &= \{CPU, RAM\} \\ A(f_2) &= \{RAM\} \\ A(CPU) = A(RAM) &= \{\} \end{aligned}$$

Using these sets and relationships, a fault tree model can be generated from the component fault tree elements and the failure dependencies that reflects the failure behavior of both architecture layers in a conservative way. For every failure dependency relation, all basic events that are included in the component fault tree of the dependency element are added to all failure modes of the dependent component.

If $c$ has a component fault tree, then it is

$$C\tilde{F}T(c) = cft, cft \neq \emptyset.$$

If $c$ has input and output failure modes, it is

$$I_{fm}(in) \neq \{\} \text{ and } O_{fm}(out) \neq \{\}$$

for an inport $in \in I(c)$ and an outport $out \in O(c)$. In the example system as depicted in figure 2, the input and output failure modes related to the ports are

$$O_{fm}(p_1) = O_{fm}(p_2) = O_{fm}(p_3) = loss\ of$$

$$I_{fm}(p_4) = I_{fm}(p_5) = I_{fm}(p_6) = loss\ of$$

If a component $f_2$ is dependent of the correct function of another component $RAM$, the failure modes of $RAM$ trigger all failure modes of $f_2$. This is a conservative assumption, which is for sure an overestimation, but simplifies the modeling of dependencies, since there is no need to map single failure modes from $RAM$ to $f_2$. Instead, the failure modes of $RAM$ is added to all failure modes of $f_2$ using an OR gate. If multiple dependency relations are present, e.g. $f_1$ is dependent from $RAM$ and $CPU$, all failure modes are included from $RAM$ and $CPU$ into the failure behavior of $f_1$. Figure 3 shows this for the example system. The grey elements are added from the hardware layer into the component fault tree model of the software layer. The triangle at the AND gate mark failure behavior that is outside the model.

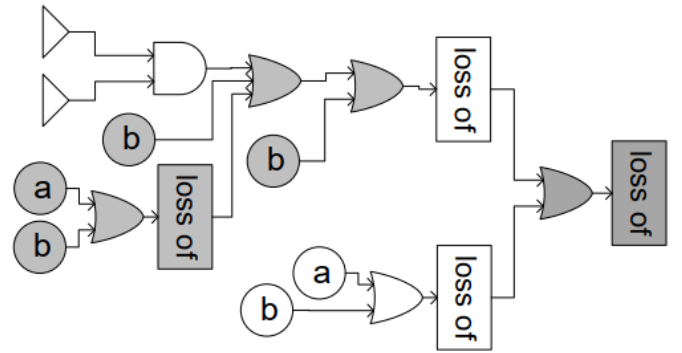

Fig. 3. Generated fault tree model from the component fault trees as depicted in figure 2 and the failure dependency relations.

Since the model as depicted in figure 3 is generated using certain rules, it can be reduced. This reduction, that simply reduces the Boolean formula, is depicted in figure 4. As can be seen in this figure, the basic events $A$ and $B$ are included in the generated fault tree. They represent the failure behavior of the hardware. Furthermore, the AND gate in the figure represents the failure behavior of the redundant sensor values, which is part of the software. In this example, two safety analysis from different layers, logical and physical, are combined for a common safety analysis model.

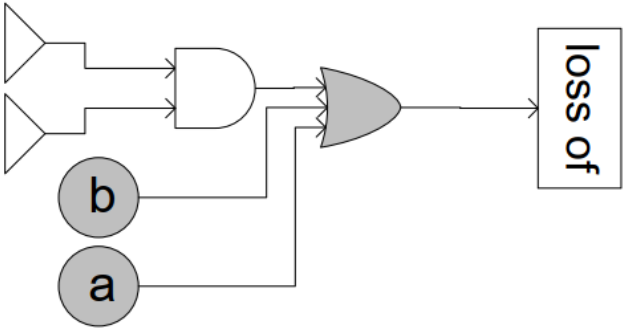

Fig. 4. Reduced fault tree from figure 3.

Using the previously described sets, the generation of component fault trees including architecture failure layer dependencies can be formalized.

Let $c$ be a component that is dependent from of the correct function of other components $c_1, \ldots, c_n$ with $A(c) = c_1, \ldots, c_n$ and $O_{fm}(c_i)$ the output failure modes of $C_i$ with

$$O_{fm}(c_i) = o_1^i, \ldots, o_m^i.$$

All output failure modes $O_{fm}(c)$ are now supplemented with the failure modes of the components in $A(c)$ to model the failure dependency in a conservative way. The output failure modes $O_{fm}(c) = o_1, \ldots, o_m$ are replaced by

$$O_{fm}(c) = \overline{o_1}, \ldots, \overline{o_m}$$

with

$$\overline{o_j} = o_j \bigvee_{i=1}^{n} o_1^i \vee \ldots \vee o_m^i.$$

In the next section, this formalized algorithm is applied within a case study.

## 5 CASE STUDY

In this section, a case study is presented from the automotive domain to demonstrate how the ALFRED approach can be used to enable the methodology of component fault trees for a layered system architecture. The presented system is a radio-controlled demonstrator vehicle as depicted in figure 6.

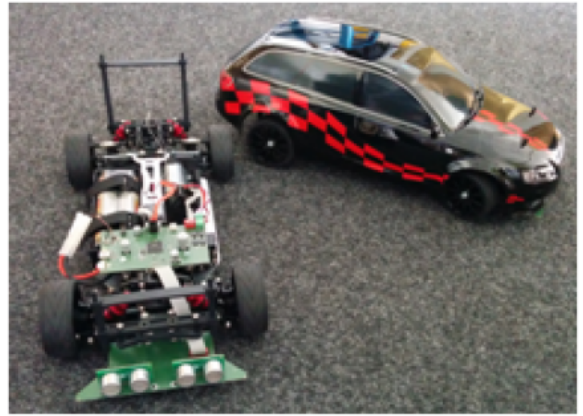

Fig. 6. Demonstrator vehicle.

Figure 5 shows the architecture of a system as a SysML internal block diagram that uses two ultra sonic sensors to enable an emergency braking functionality in a radio controlled car. The components *RadioCeceiver* (R), *UltrasonicSensor1* (U1), *UltrasonicSensor2* (U2), *EmergencyBrakingControl* (EBC), *Engine* (E) and *Steering* (S) belong to the functional layer of the architecture and exchange data via the modeled ports and interconnections. The components *Battery* (B) and *Microcontroller* (M) belong to the physical layer of the architecture.

The functionality of the emergency braking software component is basically to transmit the steering and throttle commands form the radio receiver to the steering and engine actors. If one of the ultrasonic sensors detects an obstacle, the throttle and steering signals are no longer forwarded to the actors. The car is set into emergency braking mode instead, where the actors are used to safely brake the car and omit forward moving signals for a certain time span.

The component *EmergencyBrakingControl* is a software component that *runs on* the component *Microcontroller*. The correct functionality of EBC is dependent from the correct function of M. Failures of M can influence the correct behavior of EBC functionality. Therefore EBC has an ALFRED connection to M (dotted line).

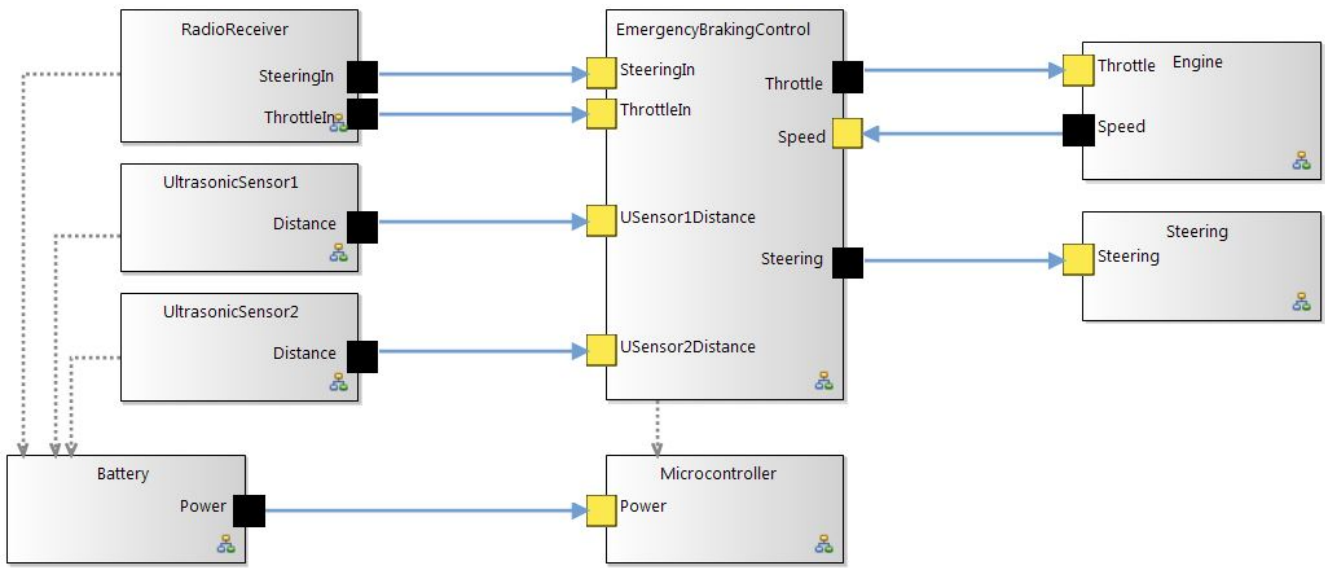

Fig. 5. Architecture system model as a SysML internal block definition diagram.

The components *RadioReceiver*, *UltrasonicSensor1* and *UltrasonicSensor2* are so-called smart sensors, which means they have both hardware and software parts. They are dependent from the battery and therefore all have an ALFRED connection to it. The components *Engine* and *Steering* are smart actors, also with both hardware and software parts. They are not dependent from the correct functionality of the micro controller. An ALFRED connection to the battery is possible, since there is only one power source for the system. But since we are only interested in the top event of loosing the emergency braking functionality and a power loss of the battery would result in a stopped car, ALFRED connections are not used here.

The comparatively simple functionality of the emergency braking functionality results in 22 different failure modes modeled within component fault trees as so-called output failure modes as introduced in section 3. The battery and the micro controller itself have no functional interaction with the other components of the system as it is modeled. Therefore, they also have no ports with the rest of the system that would allow to model the failure behavior using component fault trees. Nevertheless, their failure behavior influences the functionality.

For example, the battery is a source for failures that influence the braking functionality, since the battery powers the ultrasonic sensors.

Within the components of the functional layer, the methodology of component fault trees eases a compositional development approach. Components, for example the ultrasonic sensors, can easily be replaced by different ones. The component fault tree for such a sensor only contains the information that is related to the sensor and can be exchanged together with the component. The failure behavior of changes within one component can be expressed using the corresponding component fault tree. Including battery-related information, for example, within the component fault tree of the ultrasonic sensor would hinder an exchange using a sensor component that is modeled independently from the system. The reason is the context-related information of the battery that would be contained in such a component fault tree.

Using ALFRED connections keep the component fault trees independent from failures in different layers. Figure 7 shows the component fault tree of the component *UltrasonicSensor1*. The internal basic event labeled as *False-nagative* results in an undetected obstacle. The basic event labeled as *False-positive* results in an erroneously detected obstacle. The output failure modes of this component depicted as black triangles are used in the component fault trees of other components to model the system's top event *no emergency braking*. Due to the lack of space, all these component fault trees are not depicted here, but contain failure informations as described in section 3.

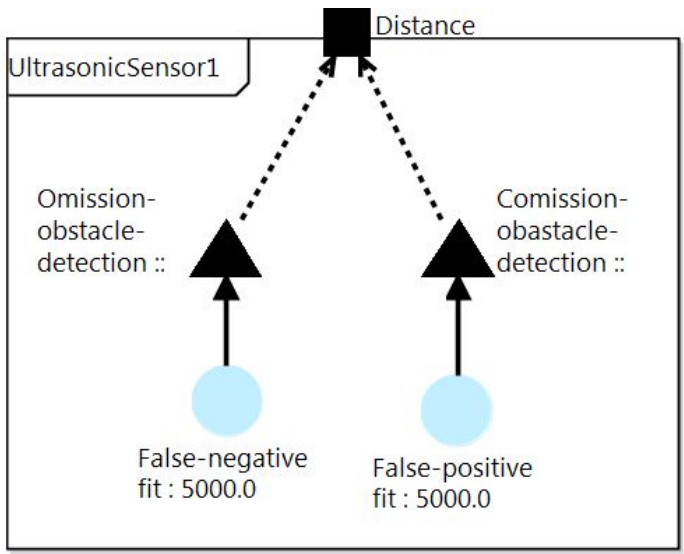

Fig. 7. Component fault tree for the component *UltrasonicSensor1*.

Applying the methodology as described in section 4 to the component fault trees and ALFRED connections in this system, allows to include the failure information of the micro controller and the battery in the fault tree analysis. The top event *no emergency braking* models the loss of the emergency braking functionality. The ALFRED methodology results in the fault tree analysis for the top event *no emergency braking* represented by the following minimal cutset analysis[1]:

E.Speed-too-low∨
EBC.HW-defect_PartCount∨
EBC.Loss-of-power ∨

(U1.Battery-too-low ∧ U2.Battery-too-low)∨
(U1.Battery-omission ∧ U2.Battery-omission)∨

(U1.Battery-omission ∧ U2.Battery-too-low)∨
(U1.Battery-too-low ∧ U2.Battery-omission)∨

---

[1]. A minimal cutset analysis is a common way to analyze fault trees. The shortest credible way through the tree from fault to initiating event is called a minimal cut set.

(U1.False-negative ∧ U2.False-negative)∨
(U1.False-negative ∧ U2.Battery-omission)∨
(U1.False-negative ∧ U2.Battery-too-low)∨
(U2.False-negative ∧ U1.Battery-omission)∨
(U2.False negative ∧ U1.Battery too low)∨

Since the battery failures of the components *UltrasonicSensor1* (U1) and *UltrasonicSensor2* (U2) rely on the same basic events (so-called *common cause failures*), a boolean solver further reduces this formula to:

E.Speed-too-low ∨
EBC.HW-defect_PartCount ∨
EBC.Loss-of-power ∨
(U1.False-negative ∧ U2.False-negative) ∨
B.Battery-omission ∨
B.Battery-too-low

As can be concluded from this result, the components of the functional layer can be maintained exchangeable from the hardware or physical layer. The ALFRED connections allowed to include the failures of the battery and the micro controller without changing the development model and add ports between the physical layer and the functional layer to model the failure behavior explicitly. Applying the ALFRED methodology is an analytical step that does not change the existing models.

## 6 CONCLUSION

This paper describes the methodology of architecture layer failure dependencies, an extension of component fault trees to maintain independency of model elements in different vertical layers of a vertically decomposed system architecture model. ALFRED connections ease the modeling of common cause failures without explicitly modeling dependencies using information flow elements such as ports. This keeps layers in the model independent from each other and so supports a compositional development approach. They can be used to reuse software on different hardware, as demonstrated in the case study, but also support arbitrary layers.

Using ALFERD connections, existing safety analysis models can be reused in a more effective way and so support an early safety assessment and head towards an automated safety qualification of future cyberphysical systems. Since we already successfully using this approach for certification tasks in the railway domain, a perspective for future work can be the further development of the approach aiming for higher precision.